\documentstyle[prb,aps,epsf,floats]{revtex}

\begin{document}

\draft

\twocolumn[\hsize\textwidth\columnwidth\hsize\csname@twocolumnfalse%
\endcsname

\title{Random and aperiodic quantum spin chains: A comparative study}

\author{Ferenc Igl\'oi$^{1,2}$, Dragi Karevski$^2$ and Heiko Rieger$^3$}
\address{
$^1$ Research Institute for Solid State Physics, 
H-1525 Budapest, P.O.Box 49, Hungary$^*$\\
$^2$ Laboratoire de Physique des Materiaux, 
Universit\'e Henri Poincar\'e,
F-54506 Vand\oe uvre les Nancy, France\\
$^3$ HLRZ, Forschungszentrum J\"ulich, 52425 J\"ulich, Germany
}

\date{June 21, 1997}

\maketitle

\begin{abstract}
  According to the Harris-Luck criterion the relevance of a
  fluctuating interaction at the critical point is connected to the
  value of the fluctuation exponent $\omega$. Here we consider
  different types of relevant fluctuations in the quantum Ising chain
  and investigate the universality class of the models. At the
  critical point the random and aperiodic systems behave similarly,
  due to the same type of extreme broad distribution of the energy
  scales at low energies. The critical exponents of some averaged
  quantities are found to be a universal function of $\omega$, but
  some others do depend on other parameters of the distribution of the
  couplings. In the off-critical region there is an important
  difference between the two systems: there are no Griffiths
  singularities in aperiodic models.
\end{abstract}


]

\newcommand{\bc}{\begin{center}}
\newcommand{\ec}{\end{center}}
\newcommand{\be}{\begin{equation}}
\newcommand{\ee}{\end{equation}}
\newcommand{\beqn}{\begin{eqnarray}}
\newcommand{\eeqn}{\end{eqnarray}}

The presence of quenched disorder has a profound effect on quantum
phase transitions, especially in one-dimensions\cite{rieger97}.  A
common feature of these one-dimensional random magnets is the extremly
broad distribution of energy scales at low energies in the vicinity of
the critical point.  This property makes it possible to use
approximate renormalization-group transformations \cite{ma} , which
becomes exact at the critical point.  Recently, Fisher \cite{fisher}
have obtained new striking results for the random transverse-field
Ising spin chain defined by the Hamiltonian:
\be
H=-\sum_l J_l \sigma_l^x \sigma_{l+1}^x-\sum_l h_l \sigma_l^z\;.
\label{hamilton}
\ee
Here the $\sigma_l^x,~\sigma_l^z$ are Pauli matrices at site $l$ and
the $J_l$ exchange couplings and the $h_l$ transverse-fields are
random variables. The random couplings (and fields) in
(\ref{hamilton}) show unbounded fluctuations, the cummulated deviation
from the average coupling $[J]_{\rm av}$ grows asymptotically with the
size $L$ as:
\be
\Delta(L)=\sum_{l=1}^L\left( J_l-[J]_{\rm av} \right) \sim L^{\omega}\;,
\label{fluctuate}
\ee
with the fluctuating or wandering exponent $\omega=\omega_{\rm rand}=1/2$.  
A similar type of unbounded fluctuations occur in aperiodic sequences,
too, which are generated through substitutional rules. For example the
Rudin-Shapiro (RS) sequence is built on four letters {\bf A},{\bf
B},{\bf C} and {\bf D} with the substitutional rule:
\be
{\bf A} \to {\bf AB}~~,~~{\bf B} \to {\bf AC}~~,~~{\bf C} \to {\bf DB}~~,~~
{\bf D} \to {\bf DC}\;,
\label{rs}
\ee
thus starting with a letter {\bf A} one proceeds as: ${\bf A} \to {\bf
AB} \to {\bf ABAC} \to {\bf ABACABDB} \to$ etc, and one may assign
different couplings to the different letters. The wandering exponent
of the sequence
\be
\omega={\ln |\Lambda_2| / \ln \Lambda_1}\;,
\label{wandexp}
\ee
which is given in terms of the leading and the next-to-leading
eigenvalues $\Lambda_{1,2}$ of the substitution matrix
\cite{queffelec}, is $\omega_{\rm RS}=1/2$, i.e.  just the same as for
the random sequence.

Then naturally the question arises, which type of critical behavior
can be found in quantum Ising chains with an unbounded aperiodic
modulation in the couplings (or fields). It is known from the
Harris-Luck relevance-irrelevance criterion \cite{harris,luck} that
the phase transition in an aperiodic (or quasi-periodic) quantum Ising
chain belongs to the Onsager universality class, only if the
fluctuations in the couplings are bounded, $\omega < 0$. For marginal
sequences with $\omega=0$ non-universal critical behavior and coupling
dependent anisotropy exponent was found in exact calculations
\cite{igloiturban96,bercheberche,grimmbaake}.

In the relevant situation with $\omega>0$ there are only a few exact
results \cite{igloiturban94}, however, on the basis of scaling
considerations \cite{luck,igloi93,igloiturban94,grimmbaake} the phase
transition in relevantly aperiodic and random systems seem to be
similar. For instance the energy gap of a finite, aperiodic chain at
the critical point scales as:
\be
\Delta E(L) \sim \exp(-{\rm const}\cdot L^{\omega})\;,
\label{energygap}
\ee
just as in the random model \cite{fisher,young,riegerigloi97}. Thus
one could ask the question, whether the wandering exponent $\omega$
alone is sufficient to characterize the universality classes of models
with fluctuating interactions; and whether or not the critical
exponents of the random and the RS chains are the same?

To answer to these questions in this Letter we are going to study
systematically the critical behavior of aperiodic quantum spin chains
with unbounded fluctuations and compare it with that of the random
system. We mainly concentrate on the RS model, but some results are
also presented for a family of relevant sequences defined on
$k$-letters ($k$-general sequence) {\bf $A_1$}, {\bf $A_2$},..., {\bf
$A_k$}, with the substitutional rules ${\bf A_i} \to {\bf
A_{i-1}A_{i+1}}$ for $i\le k/2$ and ${\bf A_i} \to {\bf
A_{i+1}A_{i-1}}$ for $i>k/2$ (one identifies ${\bf A_0}={\bf A_1}$ and
${\bf A_{k+1}}={\bf A_k}$). The wandering exponent (\ref{wandexp}) for
the $k$-general sequences, given by
\be \omega_k={\ln [2 \cos(\pi/k)] / \ln 2}\;,
\label{wandexpk}
\ee
is positive for $k \ge 4$. It is easy to see that the $k$-general
sequence for $k=2$ and $k=4$ is just the Thue-Morse and the
Rudin-Shapiro sequence, respectively, and obviously
$\omega_4=\omega_{\rm RS}=1/2$.

In the Hamiltonian (\ref{hamilton}) we have chosen homogeneous fields
$h_l=h$ and two-valued couplings: $J_l=\lambda$ for letters {$\bf
A_i$} with $i<(k+1)/2$, and $J_l=1/\lambda$ for letters with
$i>(k+1)/2$. For an odd $k$ we take $J_l=1$ for
$i=(k+1)/2$. Similarly, for the random model we take homogeneous
fields and binary distribution of the couplings: $J_l=\lambda$ and
$J_l=1/\lambda$, with the same probability. The critical point, simply
given by \cite{pfeuty,fisher}:
\be
[\ln J]_{\rm av}=[\ln h]_{\rm av}\;,
\label{critpoint}
\ee
is at $h_0=1$, both for the random and aperiodic systems.

In the following we calculate different physical quantities
(energy-gap, surface and bulk magnetization, critical magnetization
profiles, correlation length, etc.) for the RS-chain and compare with
the known results on the random chain. We also present some results
for the $k$-general chains, more details will be given in a subsequent
publication \cite{rslongpaper}. In the actual calculation, we first
transform $H$ in (\ref{hamilton}) into a fermionic model
\cite{fermion} and then use the representation in
\cite{igloiturban96}, which necessities only the diagonalization of a
$2L \times 2L$ tridiagonal matrix. The energy gap then corresponds to
the smallest positive eigenvalue of the matrix, whereas the
magnetization profiles can be obtained from the corresponding
eigenvectors. Details will be presented elsewhere \cite{rslongpaper}.

We are going to calculate the {\it average} quantities of finite
systems both for the random and aperiodic chains. For the aperiodic
systems the average is performed as follows. We generate an infinite
sequence through substitutions, cut out sequels of length $L$, which
start at all different points of the sequence and then average over
the realizations. Among these realizations there are $R(L)$ different,
which grows linearly with $L$: $R(L)=aL$. For the RS-chain the number
of different realizations is less than $16L$, and one can obtain the
{\it exact} average value by the following procedure. Consider the
RS-sequence in (\ref{rs}) generated from a letter {\bf A}, take the
first $4L$ sequels of length $L$, starting at positions $1,2,\dots,4L$
and take also their reflexion symmetric counterparts.  Then repeat the
procedure starting with a letter {\bf D}.  The averaging then should
be performed over these $16L$ realizations. The fact that the number
of different samples grows linearly makes it possible to obtain
numerically exact average results for relatively large ($L \le 512$)
aperiodic chains. On the other hand, for random chains with the binary
coupling distribution there are $2^L$ different realizations and we
performed the average over some ($N=50000$) randomly chosen samples.

\begin{figure}[hbt]
\epsfxsize=\columnwidth\epsfbox{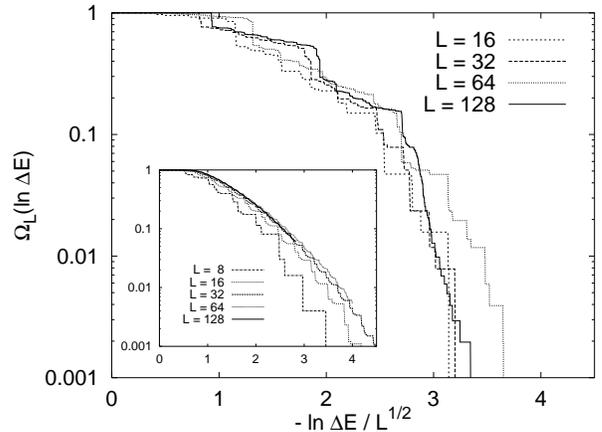}
\caption{Scaling plot of the accumulated probability density
  $\Omega(\ln\Delta E)$ versus the scaling variable
  $(\ln\Delta E/L^{1/2})$  for the RS sequence (exact average) with
  $\lambda=4$, c.f.\ eq.(\protect{\ref{accumulated}}). The insert
  shows the same for the the random sequence ($x$ and $y$ axis as in
  the main figure).}
\end{figure}

First, we investigate the probability distribution of the energy gap
$P_L(\Delta E)$ at the critical point of the RS-chain. According to
eq(\ref{energygap}) the appropriate scaling variable is $\ln \Delta E
/L^{1/2}$.  Indeed, as seen in Fig.1, the accumulated probability
distribution function
\be
\Omega_L(\ln \Delta E)=\int_{-\infty}^{\ln\Delta E} \!\!dy\,
\tilde{P}_L(y)\sim\tilde{\Omega}(\ln \Delta E /\sqrt{L})
\label{accumulated}
\ee
%
with $\tilde{P}_L(\ln\Delta E)=P_L(\Delta E)\Delta E$,
has a good data collapse using this reduced variable. Considering now
the same quantitity for the random chain one can observe the same
scaling behavior, as shown in the insert of Fig.1. Thus we can
conclude, that both the random and the RS-chains have logarithmically broad
distribution of the energy gaps at the critical point, from this fact
one expects similar consequences for the critical behavior in the two
systems.

Next we turn to consider the surface magnetization of the quantum Ising
chains, which can be obtained from the formula \cite{peschel}
\be
m_s(L,h)=\left[1+\sum_{l=1}^{L-1} \prod_{j=1}^l 
\left( {h \over J_l} \right)^2 \right]^{-1/2}\;,
\label{surfmag}
\ee
for a given realization. For one single RS-chain, starting with the letter
${\bf A}$ this expression has been evaluated exactly \cite{igloiturban94} with
the following results. For $\lambda>1$ there is a finite critical point
surface magnetization, which goes to unity as $\lambda \to \infty$, whereas for
$\lambda<1$ the surface magnetization vanishes as
\be
m_s(L,h_0) \sim \exp(-{\rm const}\cdot\sqrt{L})\;.
\label{surfmag1}
\ee
Now considering all the possible realizations of the chain at the {\it
  aperiodic} critical point eq(\ref{critpoint}) we will see, that in
{\it typical} samples i.e. which occour with probability one, the
surface magnetization vanishes as in eq(\ref{surfmag1}), whereas there
are {\it rare events} with a vanishing probability $P_{\rm rare}(L) \sim
L^{-\gamma}$, but with a surface magnetization of order unity. Thus
the critical surface magnetization is {\it non-self averaging}, it is
determined by the rare events, and its scaling dimension $x_m^s$
defined by the asymptotic relation $[m_s(L,h_0)]_{\rm av} \sim
L^{-x_m^s}$ is just $x_m^s=\gamma$.

In the following, using this observation, we calculate $x_m^s$ {\it
  exactly}.  First, we note that to each sample with a given
distribution of the couplings one can assign a walk, which starts at
the origin and takes the $l$-th step $+1$ ($-1$) for a coupling
$J_l=\lambda$ ($J_l=1/\lambda$). Then, as shown in Ref\cite{bigpaper},
taking the limit $\lambda \to \infty$ only those samples have
non-vanishing surface magnetization, where the corresponding walk
never visits sites with negative coordinates. Thus the proportion of
rare events is given by the surviving probability of the walk:
$P_{\rm rare}(L)=P_{\rm surv}(L)$. For the random chain one should consider
the random walk with $P_{\rm surv}(L) \sim L^{-1/2}$, thus one immediately
gets the exact result\cite{mccoy,fisher,bigpaper}:
\be
x_m^s({\rm rand})=1/2\;.
\label{randmag}
\ee
For the RS-chain one can perform the exact analysis\cite{rslongpaper},
from which here we present the leading finite-size dependence:
$[m_s(L,h_0)]_{\rm av}={5 \over 8} \left({1 \over \sqrt{2}} +{1 \over
    4} \right) L^{-1/2}+O(L^{-3/4})$ thus the corresponding scaling
dimension
\be
x_m^s({\rm RS})=1/2\;.
\label{rsmag1}
\ee
is the same as for the random model.

For general relevant aperiodic sequences the surface magnetization
scaling dimension can be obtained by the following scaling
consideration about the surviving probability of the corresponding
aperiodic walk. Let us perform a discrete scale transformation, which
corresponds to a substitutional step of the sequence when
the length of the walks scales as $L \to L \Lambda_1$, whereas the
transverse fluctuations as $y \to y |\Lambda_2|$. Then $N(L)$ the
number of $L$-steps surviving walks from the total of different $R(L)=aL$
walks scales as
$N(L) \to N(\Lambda_1 L)= |\Lambda_2| N(L)$, since the
number of these walks is proportional to the size of transverse
fluctuations. Consequently the surviving probability $P_{\rm
surv}(L)=N(L)/R(L)$ satisfies the scaling relation $P_{\rm
surv}(\Lambda_1 L)= |\Lambda_2|/\Lambda_1 P_{\rm surv}(L)$, with the general
result
\be
x_m^s=1-\omega\;
\label{genmag}
\ee
This simple exponent relation, which is valid for the random and RS
chains, has been numerically checked for the first members of the
$k$-general chain.

Next we turn to calculate the correlation length critical exponent $\nu$ from
the $\delta=\ln h$ dependence of the surface magnetization. In the scaling
limit $L \gg 1$, $|\delta| \ll 1$ the surface magnetization can be written as
$[m_s(L,\delta)]_{\rm av}=[m_s(L,0)]_{\rm av} \tilde{m_s}(\delta L^{1/\nu})$.
Expanding the scaling function into a Taylor series $\tilde{m_s}(z)=1+bz+O(z^2)$
one obtains for the $\delta$ correction to the surface magnetization:
\be
[m_s(L,\delta)]_{\rm av}-[m_s(L,0)]_{\rm av} \sim \delta L^{\Theta}\;.
\label{tempmag1}
\ee
with $\Theta=1/\nu-x_m^s$. This exponent can also be determined exactly in
the $\lambda \to \infty$ limit from random walk arguments \cite{bigpaper}. As
shown in Ref \cite{bigpaper} the surface magnetization of {\it rare events} is
given by:
\be
m_s(L,\delta)=(1+n)^{-1/2}-\delta {\sum_{i=1}^n l_i \over (n+1)^{3/2}}
+O(\delta^2)\;,
\label{tempcorr}
\ee
where the corresponding surviving walk returns $n$-times to its
starting point after $l_1,l_2,\dots,l_n$ steps. On the basis of
(\ref{tempcorr}) we argue that $\Theta$ is connected to the surface
fractal dimension $d_s$ of the surviving walks, defined through the
asymptotic dependence of the number of return points $n$ of a
surviving walk of $L$ steps, $n \sim L^{d_s}$. Now, to perform the
average of the linear term in (\ref{tempcorr}) one should take into
account that from the $R(L)$ different samples there are $O(n)$, which
give the most important contribution: each of those has $O(n)$ return
points of characteristic lengths $l_i \sim L$. Consequently, the
average of the linear term in (\ref{tempcorr}) grows as $L^{d_s/2}$,
thus comparing with (\ref{tempmag1}) one gets the exponent relation:
\be
{1 \over \nu}-x_m^s={d_s \over 2}\;.
\label{exprel}
\ee
The random case $\nu({\rm rand})=2$ is formally contained in
(\ref{exprel}) with $d_s=0$, since a surviving random walk returns
$n=O(1)$-times to the starting point. For the family of $k$-general
sequences $d_s=1/2$ for all values of $k$, thus the corresponding
correlation length exponent is $\nu(k)=4/(5-4\omega_k)$, with
$\omega_k$ given in (\ref{wandexpk}).  In particular for the
RS-sequence we get
\be
\nu(RS)=4/3\;,
\ee
which we also checked numerically by evaluating
eq. (\ref{tempcorr}) exactly up to $L=2^{21}$. Thus we conclude that
the {\it RS and the random Ising quantum chain are not in the same
universality class}.

Next we turn to study the magnetization in the bulk and calculate the
scaling dimension of the bulk magnetization $x_m=\beta/\nu$ from the
behavior of the average magnetization profiles at the critical point
in finite systems. As will be shown in the detailed
publication\cite{rslongpaper} the average magnetization profiles for
the RS-chain are in excellent agreement with the conformal
results\cite{burkhardt,turbanigloi97}, in a similar way as observed
for random chains\cite{profiles,conformal}. From the scaling behavior
of the profiles $x_m$ can be estimated as
\be
x_m({\rm RS})=0.160 \pm .005\;,
\label{rsmagbulk}
\ee
which also differs from the random chain value $x_m({\rm
rand})=(3-\sqrt{5})/4=.191$ \cite{fisher}.

\begin{figure}[hbt]
\epsfxsize=\columnwidth\epsfbox{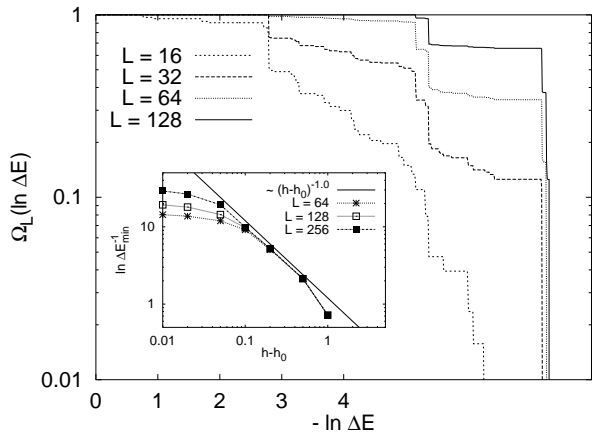}
\caption{The accumulated probability density $\Omega(\Delta E)$ for
  the RS sequence (exact average) with $\lambda=4$ slightly above the
  critical point ($h=1.5$). The distribution is chopped off at
  $\ln\Delta E_{\rm min}^{-1}(h=1.5)\approx7.3$. The insert shows
  $\ln\Delta E_{\rm min}^{-1}(h)$ versus the distance from the critical
  point ($h-h_0\sim\delta$ for $h\to h_0$) in the RS sequence for
  different system sizes. The srtraight line has slope $-1$, as
  predicted by (\protect{\ref{minigap}}).}
\end{figure}

Finally, we turn to discuss the properties of the RS-chain in the
disordered phase. For the random model, as known from exact results
\cite{mccoy,fisher}, the susceptibility diverges in a whole region, in
the so called Griffiths-phase \cite{griffiths}. This non-analytical
behavior is due to a broad distribution of the energy gaps, which
scales in finite systems as $\Delta E(L,\delta) \sim L^{-z(\delta)}$,
with the dynamical exponent $z(\delta)$ \cite{young}.  As a
consequence, in the infinite chain there is no finite time-scale and
the autocorrelation function decays algabraically with an exponent
which is related to $z(\delta)$ \cite{bigpaper}.

For relevantly aperiodic chains the same type of scenario, i.e. the
existence of a Griffiths-phase is speculated \cite{luck}. To clarify
this issue for the RS-chain we have calculated the distribution
function of the energy gap in the disordered phase. As seen on Fig.2
the accumulated distribution function for the RS-chain has a
qualitatively different behavior, than for the random chain: there is
an $L$-independent cut-off at $\Delta E_{\rm min}(\delta)$. As a
consequence in the disordered phase the autocorrelation function
always decays exponentially with a relaxation time $t_{\rm r} \sim \Delta
E_{\rm min}^{-1}(\delta)$ and the susceptibility and other physical
quantities remain analytic. Therefore there is {\it no
  Griffiths-region} in the RS-chain and a similar scenario is expected
to hold for any aperiodic quantum spin chain.

One can estimate the minimal energy gap $\Delta E_{\rm min}(\delta)$ close
to critical point from the related expression in eq(\ref{energygap})
by replacing there $L$ with the corresponding characteristic
length-scale in the off-critical region. We note, that this
length-scale is not the average correlation length, since it is
related to a single sample.  In this case the aperiodic length scale
$l_{\rm ap}$ is obtained\cite{igloi93,igloiturban94} by equating the
aperiodic (fluctuating) energy contribution $\sim l_{\rm ap}^{\omega-1}$
with the "thermal" energy $\sim \delta$ giving $l_{\rm ap} \sim
\delta^{-1/(1-\omega)}$.  Thus the minimal energy gap is expected to
scale as
\be 
\Delta E_{\rm min}(\delta)\sim \exp(-{\rm const}\cdot\delta^{-{\omega\over
    1-\omega}})\;.
\label{minigap}
\ee
This relation is indeed satisfied for the RS-chain, since in this case
according to numerical results $\ln \Delta E_{\rm min}(\delta)\sim
\delta^{-1}$, as can be seen in the insert of Fig.2.

To summarize, we have made a comparate study of the critical
properties of random and relevantly aperiodic quantum Ising chains. At
the critical point the two systems behave very similarly, some
critical exponents are found even identical. This similarity is mainly
attributed to the fact that, at the critical point the distribution
functions of different physical quantities in the two systems
qualitatively agree. Outside the critical point, where the low energy
tail of the distributions are of importance, the two systems behave
differently. In the random system, with exponentially many
realizations, there is no minimal energy scale, whereas in the
aperiodic chain with linearly many independent realizations there
exists a minimal energy cut-off. This leads then to the absence of the
Griffiths phase in aperiodic systems.

This study was partially performed during F.\ I.'s visit in Nancy. He
is indebted to L. Turban for enlightening discussions.  This work has
been supported by the French-Hungarian cooperation program "Balaton"
(Minist\`ere des Affaires Etrang\`eres-O.M.F.B), the Hungarian
National Research Fund under grants No OTKA TO12830 and OTKA TO23642
and by the Ministery of Education under grant No FKFP 0765/1997.
H.\ R.'s work was supported by the Deutsche Forschungsgemeinschaft
(DFG). The Laboratoire de Physique des Materiaux is Unit\'e de
Recherche Associ\'ee au C.N.R.S. No 155.


\begin{references}
\bibitem{conformal}
        We note that the RS quantum Ising chain, as well
        as the random chain are {\it not conformally invariant},
        since they are not isotropic. As seen from eq(\ref{energygap})
        at the critical point time- and length-scales
        are related as $\ln t_r \sim L^{\omega}$.
\end{references}
\end{document}